\documentclass{ws-procs961x669}            
\usepackage{amsmath}
\usepackage[utf8x]{inputenc}
\usepackage[dvipsnames]{xcolor}
\usepackage{physics}
\usepackage{amssymb} 
\usepackage{tensor}
\usepackage{setspace}
\usepackage{bm}
\usepackage{pzccal}
\usepackage{graphicx}
\usepackage{tikz}
\usetikzlibrary{tikzmark,fit}
\usetikzlibrary{decorations.pathreplacing}
\usepackage[makeroom]{cancel}
\usepackage{tikz-cd}

\begin{document}
\title{Symplectic evolution of an observed light bundle}

\author{N. Uzun$^*$}

\address{Univ Lyon, Ens de Lyon, Univ Lyon1, CNRS, Centre de Recherche Astrophysique de Lyon\\ 
Lyon, UMR5574, F69007, France\\
$^*$E-mail: nezihe.uzun@ens-lyon.fr\\
http://www.univ-lyon1.fr}

\begin{abstract}
Each and every observational information we obtain from the sky regarding the brightnesses, distances or image distortions resides on the deviation of a null geodesic bundle. In this talk, we present the symplectic evolution of this bundle on a reduced phase space. The resulting formalism is analogous to the one in paraxial Newtonian optics. It allows one to identify any spacetime as an optical device and distinguish its thin lens, pure magnifier and rotator components. We will discuss the fact that the distance reciprocity in relativity results from the symplectic evolution of this null bundle. Other potential applications like wavization and its importance for both electromagnetic and gravitational waves will also be summarized.
\end{abstract}


\bodymatter

\section{Introduction}\label{sec:Introduction}
When we make an observation on the sky, we deduce the brightness, distance and image distortion information about astrophysical objects via a bundle of null rays. This geodesic bundle connects the emitter and the observer that are located at two distant regions of a spacetime. Nevertheless, they receive reciprocal information about each other. 

For example, an observer located at point O on Earth, obtains the distance to a galaxy cluster located at point G, by making use of the observed solid angle at point O, and the estimated cross--sectional area on the sky at point G. This gives the angular diameter distance, $D_A$. Such a distance estimation is mostly relevant for standard rulers in cosmology as, for example, baryon acoustic oscillations. Likewise, an observer at point G can estimate the distance to a point O in a similar fashion. According to an observer on Earth, on the other hand, this information is contained in the luminosity distance, $D_L$. Namely, by using the observed flux 
\begin{eqnarray}\label{eq:Flux}
F_{\rm{observed}}= \frac{\rm{Source\,luminosity}}{\rm{Surface\,area}}=\frac{L_{\rm{source}}}{4\pi D_U^2},
\end{eqnarray}
one can deduce an \textit{uncorrected} luminosity distance, $D_U$. When the relativistic corrections are included, one obtains a corrected luminosity distance, $D_L$, via
\begin{eqnarray}
D_L=(1+z)^{-1}D_U,
\end{eqnarray}
where $z$ is the redshift factor. The relationship between the luminosity distance and the angular diameter distance is given by Etherington's distance reciprocity theorem \cite{Etherington:1933}
\begin{eqnarray}\label{eq:Etherington}
D_L=(1+z)D_A.
\end{eqnarray}

The distance reciprocity, eq~(\ref{eq:Etherington}), was originally derived for light propagation within a single spacetime. This means that the initial point of the propagation is an observation point at which the geodesic deviation vector becomes zero. Eventually, it can be shown that the distance reciprocity follows from the symmetries of the Riemann curvature tensor \cite{Etherington:1933,Ellis:1971}.

Previously, we showed that \cite{Uzun:2020} one can extend the proof for more generic scenarios. Namely, for a ray bundle propagating in multiple geometries, one needs to propagate a bundle with arbitrary initial conditions. Such a scenario was used, for example, in  Fleury \textit{et al.} \cite{Fleury:2013sna,Fleury:2014gha}. In those investigations, there exists a $4\times 4$ Wronskian matrix which takes the initial ray bundle variables to the final ones. A similar construction can be found \cite{Grasso:2019,Korzynski:2020,Korzynski:2021,Grasso:2021} where the Wronskian in question is 8--dimensional and it includes more information regarding the light bundle evolution. We demonstrated \cite{Uzun:2020} that the Wronskian matrix of Fleury \textit{et al.} is indeed symplectic and the most general form of the distance reciprocity 
follows from the symplectic symmetries of this transformation matrix. This was achieved by following a Hamiltonian formalism defined on a 4--dimensional phase space. Note that our approach is analogous to the reduced phase space optics formalism developed for classical paraxial rays \cite{Wolf:2004,Torre:2005}. 

In classical optics, one starts with Fermat's action in order to obtain a Hamiltonian formalism in which rays are the geodesic solutions of the corresponding optical metric of Fermat. Likewise, we implemented the usage of geodesic actions applied to a null bundle. The outcome is an effective geodesic deviation action which was originally derived up to higher orders in Vines' work \cite{Vines:2014}.

The aim of this talk is to give a brief summary of our previous formalism \cite{Uzun:2020}, along with discussing the importance of implementing such new techniques to gravitational optics. We summarize the aforementioned symplectic phase space approach in Section~(\ref{sec:Symplectic evolution of light bundles on phase space}). Some perspectives on the symplectic ray bundle evolution is given in Section~(\ref{sec:Some perspectives on the symplectic ray bundle evolution}) and we conclude with Section~(\ref{sec:Conclusion}).
\section{Symplectic evolution of light bundles on phase space}\label{sec:Symplectic evolution of light bundles on phase space}
\subsection{Vines' bi--local geodesic deviation action}
It is known that connecting two geodesics is, in general, non--local. Therefore, if one wants to obtain a geodesic deviation action, one needs to consider a bi--local formalism. Vines \cite{Vines:2014} achieves this for generic curves by following the definition of Synge's world function \citep{Synge:1960} $\sigma (x,y)$. This is a bi--local object which depends on two spacetime points $x$ and $y$. Those two points are connected by a unique geodesic, $\Sigma$, such that 

\[
    \sigma (x,y)= \frac{1}{2}
\begin{cases}
    (\rm{proper\,distance})^2,& \Sigma:\rm{spacelike}\\
    0,              &  \Sigma:\rm{null}\\
    -(\rm{proper\,time})^2,& \Sigma:\rm{timelike}.
\end{cases}
\]

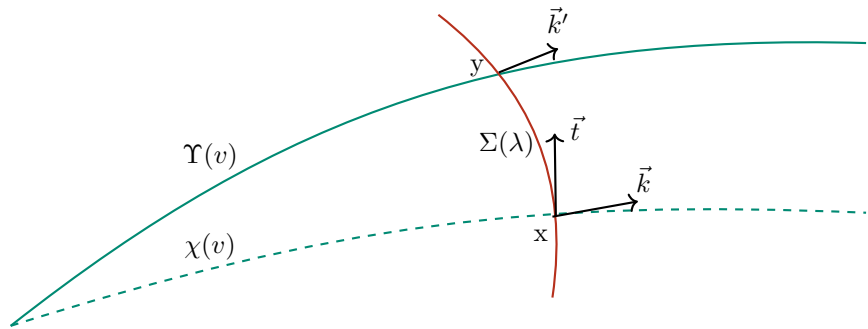
\begin{figure}
\hspace*{1cm}
\begin{center}
\begin{tikzpicture}[thick, scale=1.5]
\draw [PineGreen] (0.25,0.5) to[bend left=20] (7.75,3);
\draw [PineGreen,dashed] (0.25,0.5) to[bend left=10] (7.75,1.5);
\draw [BrickRed,thick] (5,0.75) to[bend right] (4,3.25);

\node [draw=none] at (4.9,1.3) {x};
\node [draw=none] at (4.34,2.78) {y};

\node [draw=none] at (4.6,2.1) {$\Sigma(\lambda)$};
\node [draw=none] at (2,1.2) { $\chi (v)$};
\node [draw=none] at (2,2) {$\Upsilon (v)$};

\draw [->] (5,1.465) to (5.75,1.6);
\draw [->] (5.03,1.465) to (5.02,2.2);
\draw [->] (4.53,2.74) to (5.05,2.95);

\node [draw=none] at (5.8,1.8) {$\vec{k}$};
\node [draw=none] at (5.05,3.2) {$\vec{k}'$};
\node [draw=none] at (5.2,2.25) {$\vec{t}$};
\end{tikzpicture}
\end{center}
\caption{The central null geodesic $\chi (v)$ is plotted with dashed lines. The red curve represents $\Sigma(\lambda)$ given by Synge's world function that is spacelike. The outermost null geodesic $\Upsilon (v)$ can be uniquely obtained through $\chi (v)$ and $\Sigma(\lambda)$.}\label{fig:Syngesworldfun}
\end{figure}

Now consider a bundle whose outer most null geodesic, $\Upsilon(v)$, has a tangent vector $\vec{k}'$ (See Fig~(\ref{fig:Syngesworldfun}).). The central null geodesic, $\chi(v)$, with a tangent vector $\vec{k}$, can also be parameterized with the same affine parameter $v$. This is called isosynchronous parameterization. Then those tangent vectors satisfy $\nabla_{\vec{k'}}\vec{k'}=0$ and $\nabla_{\vec{k}}\vec{k}=0$ where $\nabla$ denotes the covariant derivative operator defined at a specified point.
Moreover, physical sizes of the objects on the sky are estimated by the proper sizes. Therefore, a spacelike world function, being the measure of proper distance between two spacetime points, is the most relevant tool for our construction.

Now consider a spacelike geodesic, $\Sigma (\lambda)$, that connects $\Upsilon (v)$ to $\chi (v)$. One can define an exponential map on $\Sigma (\lambda)$ via the derivatives of the world function $\sigma (x,y)$. This results in a bi--local object $\vec{\eta}$, whose variation with respect to the affine parameter that parameterizes the null bundle, is given by \cite{Vines:2014,Uzun:2020}
\begin{eqnarray}\label{eq:Total_cov_der_xi}
\dot{\eta}^\alpha=\frac{\mathbb{D}\eta^\alpha}{dv}=\left(k^\beta\nabla_\beta+{k'}^\mu\nabla_\mu\right)\eta ^\alpha.
\end{eqnarray} 
Note that for an arbitrarily large separation one needs to consider the covariant derivatives of $\vec{\eta}$ at two distinct points along the null bundle as in eq.~(\ref{eq:Total_cov_der_xi}). In that case, one can write the action functional of $\Upsilon(v)$, i.e., 
$S_{\Upsilon}=\int \frac{1}{2}{k'}^2 dv$,
by making use of the action functional, $S_{\chi}=\int \frac{1}{2}{k}^2 dv$, of the central null geodesic, $\chi(v)$, and some additional terms \cite{Vines:2014,Uzun:2020}, i.e., 
\begin{eqnarray}\label{eq:Action_outermost}
S_{\Upsilon}=S_{\chi} + \int \frac{1}{2}\left[2\vec{k}\cdot \dot{\vec{\eta}}+\dot{\vec{\eta}}\cdot \dot{\vec{\eta}}-\tensor{R}{_{\vec{\eta}}_{\vec{k}}_{\vec{\eta}}_{\vec{k}}}+O(\vec{\eta}, \dot{\vec{\eta}})^3\right]dv.
\end{eqnarray}
Here, $\tensor{R}{^\alpha _{\gamma}_\beta_{\delta}}$ is the Riemann curvature tensor.
In order to obtain the second integral, that appears on the r.h.s of eq.~(\ref{eq:Action_outermost}), one solves eq.~(\ref{eq:Total_cov_der_xi}) for $\vec{k}'$ and substitutes it in $S_{\Upsilon}=\int \frac{1}{2}{k'}^2 dv$. However, this is not a straightforward calculation as it requires the bi--local object $\dot{\eta}^\alpha$ to be Taylor expanded in the coincidence limit. We advice the reader to see the original derivation of Vines \cite{Vines:2014} for generic scenarios or our previous paper \cite{Uzun:2020} for further details in this context.

On the other hand, for the observations on the sky, what is relevant is a thin bundle of null geodesics, rather than the ones with large separations. Thus, once the local limit of eq.~(\ref{eq:Action_outermost}) is taken for an infinitesimally thin bundle, one can consider $\vec{\eta}$ as a local geodesic deviation vector. Then, one obtains an effective action
\begin{eqnarray}\label{eq:The_action}
S=\int \left(\frac{1}{2}\dot{\vec{\eta}}\cdot \dot{\vec{\eta}}+\frac{1}{2}\tensor{R}{_{\vec{\eta}}_{\vec{k}}_{\vec{k}}_{\vec{\eta}}}\right)dv,
\end{eqnarray}
whose extremization results in a first order geodesic deviation equation
\begin{eqnarray}\label{eq:first_ord_dev}
\ddot{\eta}^\alpha=\tensor{R}{^\alpha_{\vec{k}}_{\vec{k}}_{\vec{\eta}}}\,.
\end{eqnarray}
Note that as from eq.~(\ref{eq:The_action}), and onwards, the overdot represents a local covariant derivative with respect to the central ray $\vec{k}$. 

In the next section, we will introduce a local tetrad and an associated screen basis around an observation point. This will allow us to rewrite the action functional (\ref{eq:The_action}) in terms of the observable variables in the following sections.

\subsection{Screen basis, cross--sections and solid angles} \label{sec:Screen basis, cross--sections and solid angles}
Consider an observer with 4--velocity $\vec{u}$. Assume that (s)he observes a thin null bundle with a central geodesic whose tangent vector is given by $k^\alpha = \omega \left(u^\alpha  + r^\alpha \right)$. Here, the observed frequency of light is given by $\omega=-\vec{k} \cdot \vec{u}$ and $\vec{r}$ is a spacelike vector and thus $\vec{u}\cdot \vec{r}=0$. We will also consider a 2--dimensional spacelike screen represented by the Sachs basis \cite{Sachs:1961,Perlick:2010}, $s^\alpha_{a}$, with $\{a,b\}=\{1,2\}$. Then,
\begin{eqnarray}\label{eq:Sachs_basis}
\vec{s}_{a}\cdot \vec{s}_{b}=\delta _{a b},
\qquad
\vec{u}\cdot\vec{s}_{a}=0,
\qquad
\vec{r}\cdot\vec{s}_{a}=0,
\qquad
\nabla _{\vec{k}}\vec{s}_{a}=0,
\end{eqnarray}
are satisfied. This orthogonal screen is the one on which the observables are projected. 

For an observational bundle, $\vec{k}\cdot\vec{\eta}=0$ holds initially as the null bundle converges into a vertex point where the observer is located. Note that this condition is preserved throughout the evolution of the bundle as (i)
$\nabla _{\vec{k}}\vec{\eta}=\nabla _{\vec{\eta}}\vec{k}$ holds due to the propagation vector, $\vec{k}$, and its Jacobi field, $\vec{\eta}$, being Lie dragged along each other  \cite{Wald:1984rg}, (ii) the propagation vector satisfies the null condition, (iii) the first order geodesic deviation equation, (\ref{eq:first_ord_dev}), is satisfied. Therefore, one can decompose the deviation vector as \cite{Perlick:2010,Uzun:2020}
\begin{eqnarray}
\vec{\eta}=\eta^{k}\vec{k}+\eta^{{1}}\vec{s}_{{1}}+\eta^{{2}}\vec{s}_{{2}}.
\end{eqnarray}
Note that we will denote components of $\vec{\eta}$ residing on the orthogonal screen as $\bm{\eta}:=\eta^{{1}}\vec{s}_{{1}}+\eta^{{2}}\vec{s}_{{2}}$. 

Those screen--projected components are useful in identifying observationally relevant objects. For example, the cross--sectional area, $d\mathcal{A}$, of an extended object and an observed solid angle, $d\Theta$, are respectively given by \citep{Ellis:1971}
\begin{eqnarray}\label{eq:Xsection_Sangle}
d\mathcal{A}:=\left|{\eta} ^{{1}}\wedge {\eta} ^{{2}}\right|, \qquad d\Theta:=\left|\frac{d{\eta} ^{{1}}}{d\ell}\wedge \frac{d{\eta} ^{{2}}}{d\ell}\right|.
\end{eqnarray}
where $|d\ell|=\omega dv$ is the infinitesimal proper length written in terms of the affine parameter, $v$, of the null bundle and the observed frequency of light, $\omega$. The symbol $\wedge$ represents the exterior product. 

In the next section, we will introduce a Hamiltonian formalism associated with the action functional, (\ref{eq:The_action}). This Hamiltonian will be defined on a phase space whose coordinates are constructed via the screen--projected deviation vector, $\bm{\eta}$, and its derivatives, $\bm{\dot{\eta}}$ .
\subsection{A reduced phase space and its Hamiltonian}
In general relativity, the standard way of constructing an optical phase space resides on a $3 + 1$ decomposition of a spacetime. Specifically, one chooses the phase space coordinates as the ones induced on a 3--dimensional spacelike hypersurface and the phase space momenta are chosen as the induced 3--momenta of a photon. Alternatively, we show that \cite{Uzun:2020} a 4--dimensional phase space can be effectively constructed for a null bundle  via treating the screen--projected deviation vector, as phase space Darboux coordinates,  $q ^{a}$, and its derivatives along the bundle as phase space momenta, $p _{b}$. Then a phase space vector can be defined as
\begin{eqnarray}\label{eq:Phase_coords}
\bm{\varkappa}=
\left[\begin{array}{c }
 q ^{a} \\
 p _{b} 
\end{array}\right]
=\left[\begin{array}{c }
 \eta ^{{1}} \\
 \eta ^{{2}} \\
 \dot{\eta} _{{1}} \\
 \dot{\eta} _{{2}}
\end{array}\right].
\end{eqnarray}
Note that, the overdot in eq.~(\ref{eq:Phase_coords}) now denotes a standard total derivative with respect to the evolution parameter $v$ as we are considering only the dyad components of $\vec{\eta}$.

Let us reconsider the effective geodesic deviation action functional given in eq.~(\ref{eq:The_action}). By (i) treating the integrant of the action functional as a Lagrangian, (ii) using the symmetries of the Riemann tensor and (iii) rewriting the Lagrangian in terms of the phase space coordinates, one can obtain a reduced Lagrangian. A passage to the Hamiltonian formalism gives us a reduced Hamiltonian 
\begin{eqnarray}\label{eq:red_Hamiltonian}
H=\frac{1}{2}\tensor{\delta}{^{a}^{b}} \dot{\eta}_{a}\dot{\eta}_{b}-\frac{1}{2}\tensor{\mathcal{R}}{_{\,}_{a}_{b}}\eta^{a}\eta^{b},
\end{eqnarray}
where $\tensor{\mathcal{R}}{_{\,}_{a}_{b}}:=\tensor{R}{_{a}_{\vec{k}}_{\vec{k}}_{b}}$ is known as the \textit{optical tidal matrix} in cosmology \citep{Seitz:1994, Perlick:2010}.

Then, the corresponding Hamiltonian equations can be written as a matrix equation
\begin{eqnarray}\label{eq:Ham_eqs_matrix}
\dot{\bm{\varkappa}}=\mathbf{L}_{\mathbf{{H}}}\,\bm{\varkappa},\qquad {\rm{with}} \qquad 
\mathbf{L}_{\mathbf{{H}}}=
\left[
\begin{array}{c|c}
\mathbf{0_2} & \, \, \delta ^{ab} \\
\hline
\tensor{\mathcal{R}}{_{\,}_{a}_{b}} & \, \, \mathbf{0_2}
\end{array}
\right],
\end{eqnarray}
where $\mathbf{L}_{\mathbf{{H}}}$ is the Hamiltonian matrix representation of the Lie operator defined through the reduced Hamiltonian, $H$, given in eq~(\ref{eq:red_Hamiltonian}) and $\mathbf{0_2}$ is a $2\times 2$ zero matrix. 

Let us now investigate the solutions of the Hamiltonian equations, (\ref{eq:Ham_eqs_matrix}), in the next section.
\subsection{Symplectic evolution and its symmetries}
The Hamiltonian equations, (\ref{eq:Ham_eqs_matrix}), presented in the previous section are linear equations as $\mathbf{L}_{\mathbf{{H}}}=\mathbf{L}_{\mathbf{{H}}}(v)$. Then, its solutions are given by a linear transformation of  an input phase vector, $\bm{\varkappa}_0$, defined at $v_0$, which gives an output phase space vector, $\bm{\varkappa}$, at some value of $v$, i.e.,
\begin{eqnarray}\label{eq:Transfer}
\bm{\varkappa}=\mathbf{T}\left(v,v_0\right)\bm{\varkappa}_0.
\end{eqnarray}
Substituting eq.~(\ref{eq:Transfer}) back into eq.~(\ref{eq:Ham_eqs_matrix}) shows that the ray bundle transfer matrix, $\mathbf{T}$, satisfies a similar equation as the phase space vector $\bm{\varkappa}$, i.e., $\dot{\mathbf{T}}=\mathbf{L}_{\mathbf{{H}}}\,\mathbf{T}$. Then, its solution is given by taking an ordered exponential (OE) map,
\begin{eqnarray}\label{eq:Transfer_OE}
\mathbf{T}\left(v,v_0\right)={\rm{OE}}\left[{\int _{v_0}^{v}\mathbf{L}_{\mathbf{{H}}}dv}\right]\mathbf{T}\left(v_0,v_0\right), \qquad {\rm{with}} \qquad \mathbf{T}\left(v_0,v_0\right)=\mathbf{I_4},
\end{eqnarray}
and $\mathbf{I_4}$ is a $4\times 4$ identity matrix.
In that case, $\mathbf{T}$, is a symplectic matrix which satisfies
\begin{eqnarray}\label{eq:Symplectic_T}
\mathbf{T}^{\intercal}\,\mathbf{\Omega}\,\mathbf{T}=\mathbf{\Omega}, \qquad \rm{det}\,\mathbf{T}=1,
\end{eqnarray}
as exponential map of Hamiltonian matrices are symplectic matrices. Here, $^\intercal$ denotes the transpose operator and $\mathbf{\Omega}$ is the well--known \textit{fundamental symplectic matrix} whose components are given by
\begin{eqnarray}
\Omega ^{ij}=
\left[
\begin{array}{c|c}
\mathbf{0_2} & \, \, \mathbf{I_2} \\
\hline
\mathbf{-I_2} & \, \, \mathbf{0_2}
\end{array}
\right].
\end{eqnarray}
This allows one to treat the ray bundle evolution in general relativity analogous to a dynamical problem in classical mechanics. A similar approach has indeed been developed in paraxial Newtonian optics much earlier (Cf. references in \cite{Wolf:2004,Torre:2005}). In that case, this approach is also known as the $ABCD$--\textit{matrix method} as it is very common to represent a symplectic matrix, $\mathbf{T}$, in a block form
\begin{eqnarray}\label{eq:Transfer_block_GR}
\mathbf{T}=
\left[
\begin{array}{c|c}
\mathbf{A} & \mathbf{B} \\
\hline
\mathbf{C} & \mathbf{D}
\end{array}
\right].
\end{eqnarray}
In our case, the submatrices in eq.~(\ref{eq:Transfer_block_GR}) are all $2\times 2$ matrices which satisfy certain symmetry conditions
\begin{eqnarray}\label{eq:symp_conds}
&\mathbf{A}\mathbf{B}^{\intercal},\,\mathbf{A}^{\intercal}\mathbf{\mathbf{C}},\,\mathbf{B}^{\intercal}\mathbf{\mathbf{D}}\,\,\rm{and}\,\, \mathbf{\mathbf{C}}\mathbf{\mathbf{D}}^{\intercal}\,\, \rm{are\,\,symmetric},\nonumber \\
&\mathbf{A}\mathbf{\mathbf{D}}^{\intercal}-\mathbf{B}\mathbf{\mathbf{C}}^{\intercal}=\mathbf{I_2},\label{eq:symm_matrices}
\end{eqnarray}
due to $\mathbf{T}$ being a symplectic matrix, i.e., eq.~(\ref{eq:Symplectic_T}) is satisfied. Then, the Hamiltonian equations, (\ref{eq:Ham_eqs_matrix}), become equivalent to,
\begin{eqnarray}\label{eq:four_sets}
\mathbf{\dot{A}}&=&\mathbf{C},\qquad \,\,\,\, \mathbf{A}\left(v_0,v_0\right)=\mathbf{I_2},\\ \nonumber
\mathbf{\dot{B}}&=&\mathbf{D},\qquad \,\,\,\, \mathbf{B}\left(v_0,v_0\right)=\mathbf{0_2},\\ \nonumber
\mathbf{\dot{C}}&=&\bm{\mathcal{R}}\mathbf{A},\qquad  \mathbf{C}\left(v_0,v_0\right)=\mathbf{0_2},\\ \nonumber
\mathbf{\dot{D}}&=&\bm{\mathcal{R}}\mathbf{B},\qquad  \mathbf{D}\left(v_0,v_0\right)=\mathbf{I_2}.\\ \nonumber
\end{eqnarray}
Therefore, once the solutions of the equation set above is found, one can find the explicit form of the transfer matrix, $\mathbf{T}$, and the geodesic deviation variables of the null bundle in question. This allows one to estimate certain observables as we discuss in the next section.

\section{Some perspectives on the symplectic ray bundle evolution}\label{sec:Some perspectives on the symplectic ray bundle evolution}

\subsection{Distance reciprocity}
\begin{figure}
    \centering
    \begin{tikzpicture}[thick,baseline=-0.5cm]
\draw  (0,0) -- (6,0);
\draw [BrickRed] (0,0) to[bend right] (0,1.5);
\draw [BrickRed] (0,0) to[bend left] (0,1.5);
\draw [BrickRed] (6,0) to[bend right=20] (0,1.5);
\draw [BrickRed] (5.25,0) arc (180:145:0.75);

\draw [PineGreen] (6,1) to[bend right] (6,0);
\draw [PineGreen] (6,1) to[bend left] (6,0);
\draw [PineGreen] (6,1) to[bend right=20] (0,0);
\draw [PineGreen] (0.75,0) arc (0:30:0.75);

\node [draw=none, PineGreen] at (0,-0.25) {G};
\node [draw=none, BrickRed] at (6,-0.25) {O};

\node [draw=none, PineGreen] at (1.1,0.23) {$d\Omega_G$};
\node [draw=none, BrickRed] at (4.85,0.23) {$d\Omega_O$};

\node [draw=none, PineGreen] at (6.6,0.47) {$d\mathcal{A}_O$};
\node [draw=none, BrickRed] at (-0.68,0.7) {$d\mathcal{A}_G$};


\end{tikzpicture}
    \caption{Observer at O measures a solid angle $d\Omega_O$ and estimates the cross-sectional area of an astrophysical object at G as $d\mathcal{A}_G$. Same thing is applicable for an observer at G as well. Reciprocity of the estimated distances is written as ${{d\mathcal{A}_O}/{d\Omega_G}}=\left(1+z\right)^2{{d\mathcal{A}_G}/{d\Omega_O}}$.}
    \label{fig:emitter_observer}
\end{figure}
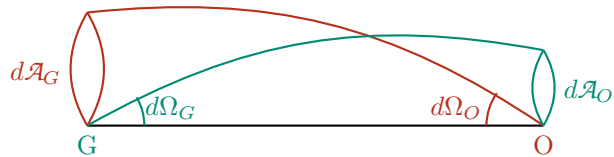

The relationship of the ray bundle transfer matrix, $\mathbf{T}$, to the observables is not immediately obvious. In order to estimate the distance to an object located at G, for example, one starts with propagating the bundle with the initial phase space vector $\bm{\varkappa}_o=(\mathbf{0},\mathbf{\dot{\eta}}_o)^\intercal$ at the observation point O (See Fig~(\ref{fig:emitter_observer}).).
Then, one can calculate the cross--sectional area, $d\mathcal{A}$, at G by (i) making use of the definitions of $d\mathcal{A}$ and the solid angle, $d\Theta$, given in eq.~(\ref{eq:Xsection_Sangle}), (ii)   inputting the information of the measured $d\Theta$ at point O in the initial phase space vector, $\bm{\varkappa}_o=(\mathbf{0},\mathbf{\dot{\eta}}_o)^\intercal$, and (iii) considering the ray bundle transformation matrix $\mathbf{T}$ that transfers the initial phase space vector to a final one as in eq.~(\ref{eq:Transfer}). In that case, the estimated distance in question is the angular diameter distance given by $D_A=(d\mathcal{A}_G/d\Theta_O)^{1/2}$. Likewise, a similar distance estimation can be done by an observer located at point G. Then, the estimated distance is the luminosity distance given by $D_L=(d\mathcal{A}_O/d\Theta_G)^{1/2}$.

As the solution of a phase space vector is given by eq.~(\ref{eq:Transfer}) in which $\mathbf{T}$ is represented in a block form (\ref{eq:Transfer_block_GR}), one obtains the angular diameter and luminosity distances respectively by making use of only the submatrix $\mathbf{B}$, i.e.,
\begin{eqnarray}\label{eq:D_A_D_L_B}
{D}_A=\omega _O{\rm{det}}\left[\mathbf{B}\left(v_G,v_O\right)\right]^{1/2},\qquad 
{D}_L=\omega _G{\rm{det}}\left[\mathbf{B}\left(v_O,v_G\right)\right]^{1/2}.
\end{eqnarray}
Here, $\omega _O$ and $\omega _G$ are the frequencies measured at respective points through which a redshift parameter is defined as $1+z=\omega _G/\omega _O$. Note that the frequency terms appear in the process of transforming the proper length, as it appears in the definition of solid angle in eq.(\ref{eq:Xsection_Sangle}), to the affine parameter, $v$, as we discussed in Section~(\ref{sec:Screen basis, cross--sections and solid angles}). The abbreviation ``${\rm{det}}$'' refers to the determinant of a given matrix.

In order for Etherington's distance reciprocity, eq.~(\ref{eq:Etherington}), to hold one needs to show that ${\rm{det}}\left[\mathbf{B}\left(v_G,v_O\right)\right]={\rm{det}}\left[\mathbf{B}\left(v_O,v_G\right)\right]$ holds. Indeed, this was previously proven by making use of the symmetries of the Riemann tensor \cite{Perlick:2010}. Note that such a case is applicable for light propagation within a single geometry. For scenarios that include light propagation in multiple spacetimes, one needs to propagate the bundle not only from the initial vertex point but also from arbitrary initial phase space points to other arbitrary ones. In that case, the information about the light bundle transformation is not only contained in the submatrix $\mathbf{B}$ but in all of the submatrices of $\mathbf{T}$. For example, this is applicable for light propagation in Swiss--cheese type models \cite{Fleury:2013sna,Fleury:2014gha}. 

Previously, we showed that distance reciprocity follows from the symplectic conditions, eq.~(\ref{eq:symp_conds}), of the ray bundle transfer matrix for generic scenarios \cite{Uzun:2020}. To be more specific, consider a ray bundle matrix, $\mathbf{T}(v_f,v_i)$, which takes an initial phase space vector, $\bm{\varkappa}(v_i)$, to a final one, $\bm{\varkappa}(v_f)$. Such a linear transformation should be equal to its inverse, once the initial and final points are traversed, i.e., $\mathbf{T}(v_i,v_f)=\mathbf{T}^{-1}(v_f,v_i)$. Moreoever, symplectic matrices form a group, thus inverse of a symplectic matrix is also symplectic. This requires $\mathbf{T}^{-1}(v_f,v_i)=\mathbf{\Omega}^{-1}\,\mathbf{T}^{\intercal}(v_f,v_i)\mathbf{\Omega}$. Then, it is easy to show that the distance reciprocity indeed follows from $\mathbf{T}(v_i,v_f)=\mathbf{\Omega}^{-1}\,\mathbf{T}^{\intercal}(v_f,v_i)\mathbf{\Omega}$. See our previous work \cite{Uzun:2020} for further details in which we show this result more explicitly.
\subsection{Decomposition of an optical propagation}
One of the earliest predictions of Einstein's general relativity was on the light propagation within our Solar System. Back then,  the idea of a spacetime structure was so hard to grasp that the astronomers conceptualized the effect of a massive object on the trajectory of light as ``bending''. Similarly, the name ``gravitational lensing'' remained within the community after Eddington performed his famous Solar eclipse experiment. Even more than 100 years after this observation, the researchers in the field use the term gravitational lensing in a loose sense.

On the other hand, in Newtonian optics a lens is defined more rigorously. For the paraxial case, for example, it is identified as a linear transformation that creates a shearing affect on the phase space defining the ray propagation. Though, lenses are not the only type of optical devices and there is a common practice in Newtonian optics in order to identify the components of light propagation. This is achieved by an Iwasawa decomposition \cite{Iwasawa:1949} where a symplectic ray transformation matrix is decomposed into its submatrices belonging to a nilpotent subgroup,  an abelian subgroup and a maximally compact subgroup for 1--dimensional systems. This allows one to identify the thin lens, the pure magnifier and the fractional Fourier transformer components of an optical propagation problem uniquely. In higher dimensions, such an identification is known as Iwasawa factorization whose components can be obtained via the submatrices of the transformation matrix $\mathbf{T}$ (Cf. \cite{Wolf:2004} for further details). 

Previously, we proposed that any spacetime can be treated as an optical device within our formalism \cite{Uzun:2020} which allows one to identify the lenses, the magnifiers and the rotators of an optical propagation in general relativity in a rigorous manner. In that case, spacetimes can be compared and contrasted in terms of their significance in different types of optical transformations, by making use of the Iwasawa factorization of their symplectic ray bundle transfer matrices. 

\subsection{From rays back to waves}
Finding the exact solutions of the Maxwell's equation on a curved background is not an easy task in general relativity. There exists a similar problem in Newtonian optics for light propagation in media with an arbitrary refractive index. Within the paraxial regime of Newtonian optics, there exist certain ``wavization'' techniques that are adopted from quantum mechanics in order to find approximate solutions of the Maxwell equations.

The analogy between classical optics and quantum mechanics follows from the similarities between the Schr{\"{o}}dinger equation of a particle wavefunction  and the Maxwell equation of a complex amplitude of a classical wave in the scalar theory \cite{Dragoman:2002,Dragoman:2005}. The $\hbar \rightarrow 0$ limit of quantum mechanics is said to give classical mechanics, and $\lambda \rightarrow 0$ limit of wave optics gives ray optics. Moreover, both the classical mechanics and the paraxial ray optics are constructed on symplectic phase spaces. Thus, in the first order limit, a wave picture can be attained back by making use of some phase space techniques that were initially introduced within quantum mechanics. In that case, phase space distribution functions, like Wigner function (or other analogous definitions), and metaplectic operators that are responsible for the spreading of a wave, play the central role in the wavization process \cite{Torre:2005}. 

We proposed \cite{Uzun:2020} that such methods can be adopted from Newtonian optics as we also reduced the problem of light bundle propagation into a symplectic evolution problem in phase space. Note that this is immediately relevant for gravitational lensing studies because when the wavelength of light is not much smaller than the gravitational lensing object, the wave effects on lensing become important. It is known that the wave effects amplify the estimated intensity of a wave in general \cite{Nakamura:1999,Matsunaga:2006uc}. Also, an earlier result showed that the wave and ray optics do not agree in terms of the position of the image formation for a specific geometry \cite{Herlt:1978}. However, there is not much known in terms of the image formation in wave optics of light \cite{Nambu:2013}. There are limited number of studies that investigate the wave optics and those usually only focus on the lensing problem on static background \cite{Nambu:2016,Turyshev:2017,Turyshev:2018}. We propose that symplectic methods can be quite useful in this respect.

Moreover, it is known that gravitational waves also have null propagation vectors similar to light rays. Therefore, geometric optics is relevant for the investigation of gravitational waves at the linear order \cite{Thorne:1983,Dolan:2018}. Also, the interaction of gravitational waves with matter fields is known to be weak. Thus, their coherence is expected to be preserved even over cosmological distances  \cite{Baraldo:1999}. Then, the interplay between the ray and wave behaviours can be used to study gravitational waves within our formalism. Also note that wave effects on the lensing of the gravitational waves have been investigated in some studies only lately \cite{Takahashi:2003ix,Moylan:2007fi,Nambu:2013,Yoo:2013cia,Takahashi:2016jom,Dai:2017huk,Dai:2018,Meena:2019ate}. However, all of those studies focus on wave effects on specific geometries. 

In addition, recently, new ideas are introduced to the literature regarding the wave effects. For example, the spin Hall effect which is normally studied in condensed matter field, is known to bring corrections to the geometrical optics limit in the case of light rays. Oancea \textit{et al.} \cite{Oancea:2020} found an analogous effect for the gravitational optics and Andersson \textit{et al.} \cite{Andersson:2020gsj} found a similar effect for the gravitational waves. Similarly, the Aharonov-Bohm effect is known from quantum electromagnetism. Baraldo \textit{et al.} \cite{Baraldo:1999} found that there exists an analogous effect for the gravitational waves. Namely, a shift is predicted on the interference pattern of the gravitational waves due to the angular momentum of a lensing object.

Therefore, we propose that once a wavization procedure is applied to our symplectic null bundle propagation, evolution of gravitational waves and their lensing can be studied within more generic scenarios as well. 

\section{Conclusion}\label{sec:Conclusion}
In this talk, we summarized our work \cite{Uzun:2020} which reduces an observational light bundle evolution in general relativity analogous to a classical optical problem in phase space. This requires simultaneous implementation of a null geodesic action to the outermost null curve and the central null curve of an observational light bundle. One then obtains an effective geodesic deviation action which can be written in terms of the screen--projected variables of the null geodesic deviation. This reduces the problem by one order and a 4--dimensional reduced phase space can be obtained by treating the screen--projections of the geodesic deviation vectors as canonical coordinates and their derivatives along the central null ray as canonical momenta. The Hamiltonian equations of this first order problem can be written as a matrix equation whose solutions are obtained via linear symplectic transformations. We shortly discussed (i) a generic proof of the distance reciprocity in general relativity by making use of the symmetries of the symplectic ray bundle transfer matrix, (ii) decomposition of the light bundle propagation into thin lens, pure magnifier and fractional Fourier transformation portions such that any spacetime can be rigorously identified as an optical device, (iii) the relevance of our method to the wave optics for electromagnetic and gravitational waves which have been drawing increasing interest in the literature lately.

\section*{Acknowledgements}
The author thanks Thomas Buchert for his comments. This work is a part of a project that has received funding from the European Research Council (ERC) under the European Union's Horizon 2020 research and innovation programme (grant agreement ERC advanced Grant 740021ARTHUS, PI: Thomas Buchert).
\bibliographystyle{ws-procs961x669}
\bibliography{references}

\end{document}